**Studies of protein adsorption on implant materials in relation to biofilm formation**

**I. Activity of Pseudomonas aeruginosa on Polypropylene and High density Polyethylene in presence of serum albumin.**

Authors:

*S Dutta Sinha*[1], Susmita Chatterjee[3], P.K.Maity[3], S.Tarafdar[1], S.P.Moulik[2]*

**Abstract:**

The surface of biomaterials used as implants are highly susceptible to bacterial colonization and subsequent infection. The amount of protein adsorption on biomaterials, among other factors, can affect the nature and quality of biofilms formed on them. The variation in the adsorption time of the protein on the biomaterial surface produces a phenotypic change in the bacteria by alteration of the production of EPS (exoplysaccharide) matrix. Knowledge of the effects of protein adsorption on implant infection will be very useful in understanding the chemistry of the biomaterial surfaces, which can deter the formation of biofilms. It is observed that the adsorption of BSA on the biomaterial surfaces increases with time and concentration, irrespective of their type and the nature of the EPS matrix of the bacterial biofilm is dependent on the amount of protein adsorbed on the biomaterial surface. The adsorption of protein (BSA) on the biomaterials, polypropylene (PP) and high density polyethylene (HDPE) has been studied and the formation of the biofilms of Pseudomonas aeruginosa on them has been examined.

**1. Introduction**

Infections associated with implanted devices are notoriously difficult to treat and often lead to serious complications.[1, 2, 3] As has been widely discussed,[4,5,6] the core issue of implant infection involves bacterial attachment and colonization on solid surfaces, which can be otherwise termed as **'biofilm formation'**. Biofilm has been metaphorically called a "city of microbes", [7] while the EPS matrix represent the "house of biofilm cells". EPS matrices are biopolymers originating from the bacteria in which the microbes are embedded.[8] The bacterial attachment to a biomaterial is a sort of 'bioadhesion', [9] which involves a complex interaction between the bacteria, the biomaterial surface and the liquid phase (blood, serum, plasma, urine etc.) in which the biomaterial is immersed. Protein molecules from the liquid phase get adsorbed on the surface of the biomaterial within a very short time (< 1second) after the introduction of the implant inside the body, [10] much before the arrival of microbes at the implant site. The 'arrived' microbes at the implant site, encounter a '**substrate**' which comprises biomaterial surface coated with a protein. This is followed by a 'race for the surface', [11] between the tissue cells and microbes and colonization depends upon whoever wins the race.

Interaction of proteins with material surfaces is a well known but rather complicated phenomenon. [12] One of the remarkable consequences of this interaction is that surfaces of biomaterials coated with proteins display properties which differ from the bulk of the biomaterial itself. [13] The present study shows that modulation of surface properties on the amount of protein adsorption can alter the architecture of the biofilm that forms on it. Hence, an in-depth understanding of the phenomenon of protein adsorption on a biomaterial surface is critical for developing and designing suitable biomaterial surfaces to deter the biofilm formation of bacteria. The biomaterials (polypropylene and high density polyethylene) and the protein bovine serum albumin (BSA) are used in this study, and the biofilm formation by the model bacterium Pseudomonas aeruginosa on BSA covered biomaterial surfaces is examined.

## 2. Experimental Section

### 2.1 Materials and Methods

We have taken samples of polypropylene (PP) and high density polyethylene (HDPE) which are clinically used in biomaterial implants obtained from Plastic Abhiyanta Ltd, India. PP is used in venous catheters and HDPE in orthopaedic implants. The water used in all our experiments is of HPLC grade (Lichrosolv) obtained from Merck,India. Tris buffer from Sigma Aldrich,USA is used for preparing aqueous solution of BSA obtained from MP Biomedicals Ltd, USA.

The present study consists of two parts: (1) experiment on adsorption of protein on biomaterial surfaces, (2) comparison of the biofilm growth on bare biomaterial surfaces and those with adsorbed BSA on them.

### 2.2 Experiment of adsorption

Polymer chips of PP and HDPE were obtained in square configuration (10mmx10mm).They were initially rinsed with water, blow dried and cleaned in an ultrasonic cleaner. BSA was mixed in different proportions with buffer solutions of pH 7.4 (concentration ranging from 0.5 mg/ml to 1.5mg/ml) and left for about a week with intermittent mixing to dissolve the BSA completely. BSA solutions of a specific concentration were taken in six separate glass vials each one containing a single chip for 9hrs, 12hrs, 15hrs, 18hrs, 21 hrs and 24hrs. These times are termed as **exposure time** '$\tau$' whose maximum ($\tau^{max}$) and minimum ($\tau^{min}$) were 24 and 9hrs respectively.

After the stipulated time, the chips were removed washed and rinsed with water, and finally dried and preserved in a desiccator ready for growing biofilms. The BSA solutions obtained after the removal of the chips in each case were preserved for absorbance measurements. The UV absorbance in each case was measured at room temperature in a Shimadzu 2550 UV/VIS Spectrophotometer (Shimadzu, Japan) in matched 3.0 cm quartz cells.

**2.3 Bacteria for production of biofilms**

Pseudomonas aeruginosa was used as a model bacteria. Strain used: Pseudomonas aeruginosa UC10 (wild type) collected from uro-catheter from Department of Urology, Institute of Post Graduate Medical Education & Research, Kolkata(India). It is known to be a strong biofilm former by 96 well micro-titre plate assays in a study in the Department of Microbiology, Institute of Post Graduate Medical Education & Research. [14] Medium used for culture was Luria Bertanii Agar (LBB)[ Himedia, India]

**2.4 Biofilm formation on the chips**

The polymer chips obtained from adsorption experiments were sterilized by plasma treatment to prevent the degeneration of the polymer surface as well as the protein adsorbed on the surface. They were then used for the growth of the biofilms on them.

The bacteria grown overnight in LBB at 37° C were diluted in the same medium to an optical density of 0.5 at 600 nm. The diluted culture was poured over the surface of the treated chips placed in the wells of 24 well tissue culture plates (Tarsons, India). The sterile LBB was added and incubated at 37°C for 7 days with shaking. The ratio of the culture and the sterile broth was 1:100. A positive and a negative control were included with untreated chip surface inoculated with bacteria serving as the positive control, and the untreated chips in the absence of bacteria served as the negative control. Broth was added from time to time in the wells to avoid desiccation. The chips were aseptically removed and washed with phosphate buffered saline (PBS pH 7.2). This step eliminated all the free floating bacteria and only the sessile forms remained attached to the surface. Chips were then air dried and prepared for FESEM measurements.

**2.5 Sample preparation for SEM**

The chips with attached bacterial cells were covered with 2.5% glutaraldehyde and kept for 3hours in 4°C after which they were washed thrice with the phosphate buffer solution. They were then passed once through the graded series of 25, 50 and 75 % ethanol and twice through 100% ethanol each for ten minutes. They were finally transferred to the critical point drier and kept overnight to make them ready for FESEM measurements.

To compare the EPS matrix of the biofilms produced by the Pseudomonas aeruginosa in the different conditions after 10 days, FESEM measurements of the polymer chips with and without the biofilms were taken at 5.0 kV-10 kV in a field emission scanning electron microscope (FESEM: Inspect F50, FEI Europe BV, and The Netherlands; FP 2031/12, SE Detector R580). For this purpose the dried polymer chips with and without biofilms were sputter-coated with a 3-nm thick conductive layer of gold.

# 3. Results and Discussion

## 3.1 Adsorption experiments

The BSA solution obtained from each vial was taken in a quartz cuvette and tested for absorbance in the UV- Visible spectrophotometer. For all solutions, peaks were obtained at 270 to 280 nm with varying intensities.

For a particular polymer (say PP, **Figure 1**) for a fixed exposure time, absorbance was higher for higher initial [BSA].Higher exposure time had a decreasing effect on the absorbance at each initial concentration. Thus at each [BSA], higher exposure time produced more adsorption of the protein on to the surface of the polymer chips. Thus the adsorption of

BSA was proportional to [BSA] in solution for the same exposure time. It was lower for the lesser $\tau^{min}$ for the same initial [BSA] in solution.

Comparison between the absorbance peaks of PP and HDPE (Figure 2) having the same adsorption time both 9 and 24 hours and equal [BSA] showed that the peaks of the latter were higher compared to that of the former. Thus, under similar conditions the adsorption of BSA was higher on the surfaces of PP chips compared to HDPE. Also the amount of adsorption on both PP and HDPE surfaces was more for greater exposure time for the same concentration of BSA solution.

This phenomenon may be explained in terms of hydrophobicities of the respective biomaterials and the strength of the non-covalent interactions between the biomaterial surfaces and the adsorbed protein layer. PP is more hydrophobic compared to HDPE, and the soft protein BSA prefers the PP surface more than HDPE which may be supported from the perspective of critical surface tension. PP has higher water contact angle compared to HDPE, but its critical surface tension is lower compared to HDPE allowing greater affinity and hence adsorption of BSA on it. In general, hydrophobic interaction significantly contributes to protein adsorption on a substrate. By modifying the hydrophobic polystyrene latex surface to hydrophilic, the amount of BSA adsorbed was found to decrease, suggesting the importance of hydrophobicity on protein adsorption.[15] In this context the review article of Nakanishi et al. on complicated protein adsorption on solid surfaces may be cited. [16]

**3.2 Biofilm experiments**

The FESEM photos of the biofilms on the bare PP surface and the same surface after BSA adsorbed for 9hrs and 24hrs are shown in Figure 3. Although albumin is known to prefer more tissue adhesion and less bacterial adhesion, EPS matrix in Figure 3(C) appears to be dense and hence the attachment of the biofilm to the polymer surface appears quite high as evident from the micrographs.

The SEM photos of ten day old P aeruginosa biofilms on bare HDPE surface, and the same surface adsorbed with BSA for 9hrs and 24 hrs (Figure 4), respectively reveal that the quantity and nature of EPS matrix produced by the bacteria is altered by the nature of surface encountered by it during biofilm formation, as all other factors such as temperature, pH etc., remaining the same. This also reveals that the surface encountered by the bacteria can affect their phenotype, and hence alter their genetic expression. . All the three surfaces of HDPE showed formation of more biofilms on them in comparison with PP but their extents followed the same trend in both i.e.,(biofilm on bare surface) < (biofilm on BSA covered surface).

The biofilms formed in the different instances, demonstrate that bacteria react physiologically to variations in the surface properties of the 'substrates' (surface with and without protein layer) by regulating the nature and quantity of the extracellular polymeric substances secreted by it, which in turn affects the attachment process to the biomaterial.[11] The macromolecular components of bacterial surfaces, e.g. lipopolysaccharide, protein and exopolymers, have been shown to vary in quantity and composition with the variation in the surface properties of the 'substrate'. This is again not the sole contribution of the protein or the biomaterial surface, but a combined effect of both as evidenced from both Figure 3(B) and 4(B). The adsorbed protein and the exposure time though same in both cases, the difference in the surface property of the **'substrate'** (surface with BSA adsorbed) produces a distinctly different EPS matrix in each case which has completely different consistency, making a difference in bacterial attachment of the same strain to the biomaterial surfaces. The physiological appearance of the bacteria forming the biofilm also gets altered along with the 'substrate' properties as found from Figure 3 and 4 at equal magnification (x10000).

We thus found that the phenotypic changes brought about in the bacterial cells, by encountering different **'substrates'** resulted in a difference in the ability of secretion of polymeric substances by the same strain of the bacteria. This also affects the levels of attachment of the biofilm to the biomaterial. Hence it is evident that an increase in exposure time of the 'substrate' to a protein present in blood, serum or urine will consequently produce a variation in the quantity and the nature of the EPS. A dense biofilm shown in Fig 4(C) formed on HDPE (with 24 hrs of exposure to BSA) brings forth the vulnerability of infection in orthopaedic implants, where HDPE is used (grossly due to its excellent mechanical strength). Apart from the infection standpoint, the thicker the consistency of the EPS of the biofilm, greater is the chance of thrombus formation in the blood vessels, after the maturation and detachment of the biofilm from the implant surface.

**Conclusion**

An understanding of the fundamental processes that govern protein adsorption and interaction, [17] with biomaterial surface is crucial in the endeavour to curtail implant infections. It is evident that the bacterial receptors are able to identify the differences in the substrate properties and modulate their genetic expressions accordingly. Understanding of the interaction between bacteria and biomaterial surfaces on the one hand and the interaction between proteins and biomaterials on the other would enrich our knowledge on the mechanism of prevention of biofilm formation on implants. The practical manifestation may be different for different species of bacteria, but through sustained strategic research it is quite possible to create biomaterial surfaces which will be able to modulate the genetic expression of bacteria to a non-biofilm mode even after protein adsorption.

**Acknowledgement**

This work was supported by the DST, Government of India through the Women's Scientist Scheme –A, project no. LS-466/WOS A/2012-2013.The biofilm experiments were carried out IPGMER, SSKM Hospital, Kolkata, India. Dr. S Dutta Sinha is thankful to Prof. Dr. Pratip Kundu, of School of Tropical Medicine, Kolkata, for his helpful suggestions and support. We are also thankful to Abiral Tamang and his team of researchers, comprising Tajkera, Rituparna and Sanchari at the Department of Physics (Jadavpur University) who had conducted the FESEM with utmost sincerity. Prof. S.P. Moulik acknowledges his appreciation to Jadavpur University and INSA, New Delhi for infrastructural and other supports.

Key words: protein, biofilm, biomaterial, adsorption

**References**

[1] Bisno, A. L., and F. A. Waldvogel. 1989. Introduction, p. 1-2. In A. L. Bison and F. A. Waldvogel (ed.), Infections associated with indwelling medical devices. American Society for Microbiology, Washington, D.C.

[2] Dickinson, G. M., and A. L. Bisno. 1989. Antimicrob. Agents Chemother. 33:602-607.

[3]Gristina, A. G. 1987. Science 237:1588-1595.

[4] Costerton, JW, Stewart, PS , Greenberg EP ,- 1999, Science, - sciencemag.org

[5] Hall-Stoodley, Luanne, Costerton, J. William,  & Stoodley, Paul, *Nature Reviews Microbiology* 2, 95-108 (February 2004) | doi:10.1038/nrmicro821

[6] James , G A., Swogger E , Wolcott R,  Pulcini E L,  Secor, P ,  Sestrich J ,  Costerton J W.and Stewart,  P S,. 2008, Wound Repair and Regeneration,Volume 16, Issue 1, pages 37–44,  DOI: 10.1111/j.1524-475X.2007.00321.x

[7] Watnick, P., and R.Kolter, 2000.,*J.Bacteriol*.182: 2675-2679.

[8] Flemming, H.-C., Thomas R. Neu, and Daniel J. Wozniak, *J. Bacteriol*. November 2007 vol. 189 no. 22 7945-7947.

[9] Manuel L.B. Palacio and Bharatbhushan, Phil. Trans. R. Soc. A 2012, **370**.

[10] Jeffrey J Gray, Current Opinion in Structural Biology,2004, 14:110-115. DOI. 10.1016/j.sbi.2003.12.001

[11] Subbiahdoss, Guruprakash and Kuijer, Roel and Grijpma, Dirk W. and Mei van der, Henny C. and Busscher, Henk J. *Acta Biomaterialia*, 5 (5). pp. 1399-1404. ISSN 1742-7061

[12] Nakanishi K, Sakiyama T, Imamura K. J.Biosci. Bioeng. 2001; 91: 233-244.

[13] Andrade,J.D.,Hlady, V. & Wei, A.P.Pure Appl. Chem. 64, 1777-1781.(doi:10.1351/pac199264111777).

[14] Chatterjee S., Maiti PK., Dey R., Kundu AK., Dey RK.Biofilms on indwelling urologic devices: Microbes and antimicrobial management prospects. Annals of Medical & Health Sciences Research; (2014); 4(1) 100-104.

[15] Parks, G. A. and de Bruyn, P. L., The zero point of charge of oxides. J. Phys. Chem., 66(1962) 967 – 973.

[16] K. Nakanishi, T. Sakiyama and K. Imamura, J. Biosci. Bioeng. 91 (2001) 233-244.

[17] Fattinger, C. Focal Molography: Coherent Microscopic Detection of Biomolecular Interaction. Phys. Rev. X **4**, 031024.

**FIGURES & TOC**

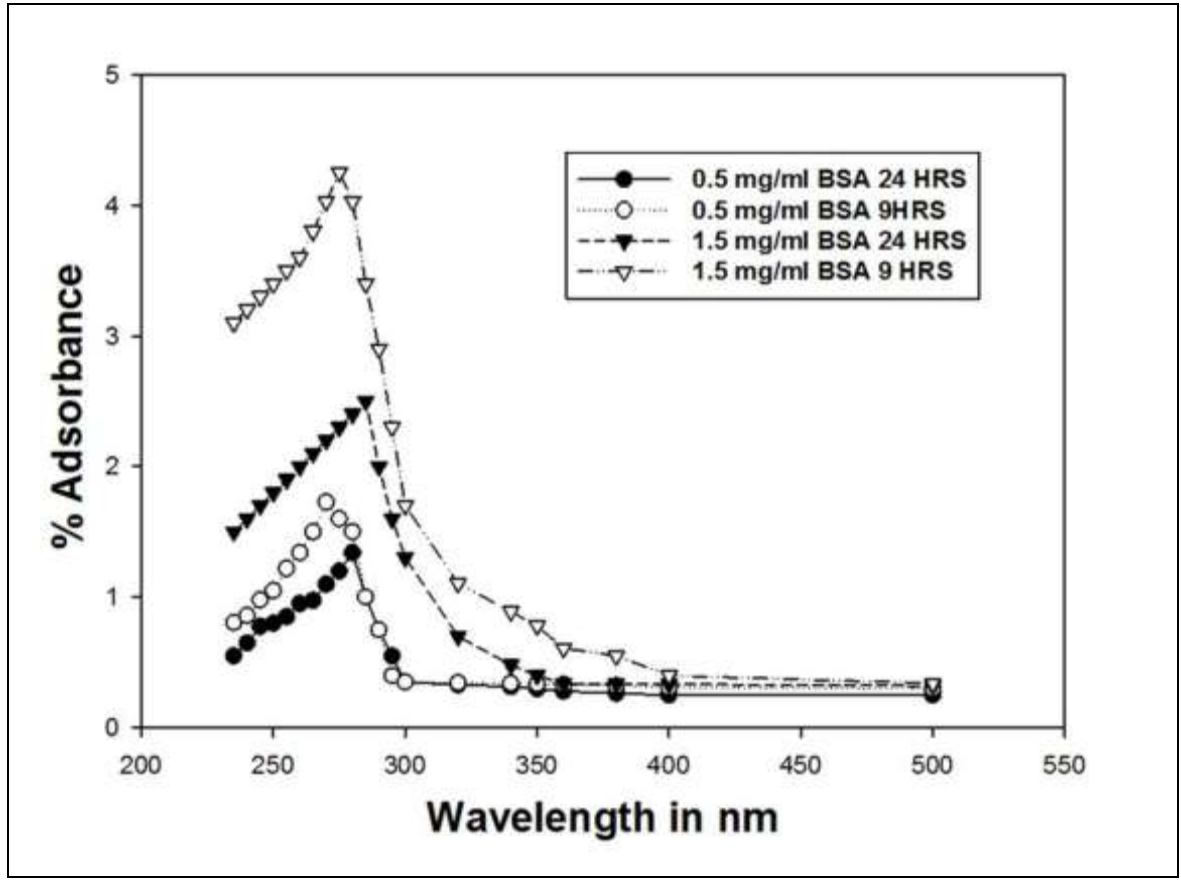

Figure 1: Absorption-wavelength profiles of BSA of varied concentrations on PP after 9 and 24 hrs.

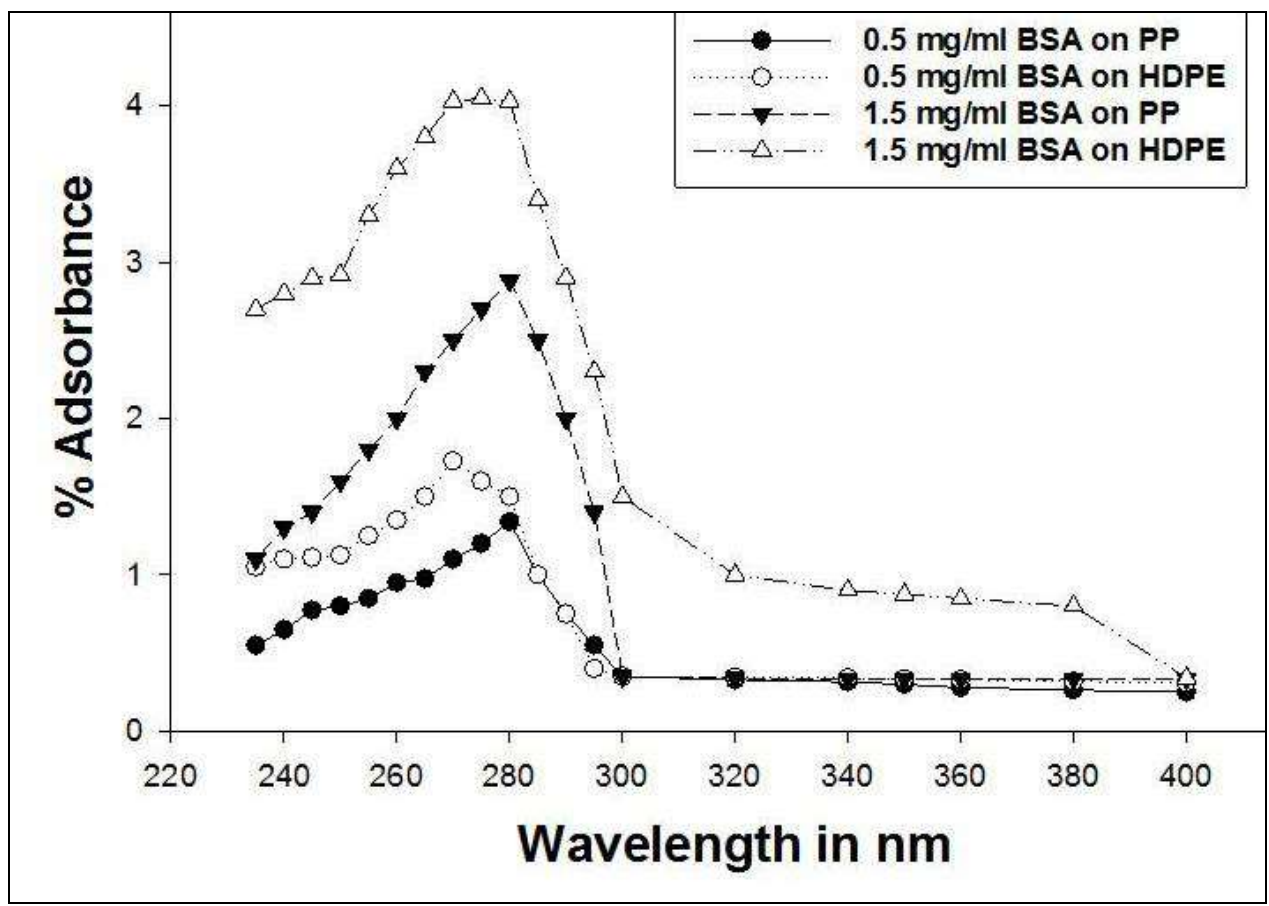

Figure 2 Comparison of % absorbance of BSA on PP and HDPE

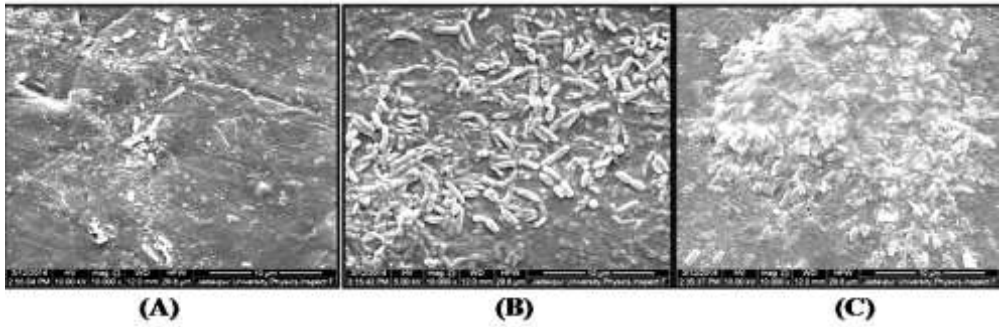

Figure 3: FESEM photographs of ten day old biofilms of P aeruginosa on PP surface

(A) Bare surface, (B) with BSA adsorbed for 9hrs, (C) with BSA adsorbed for 24 hrs.

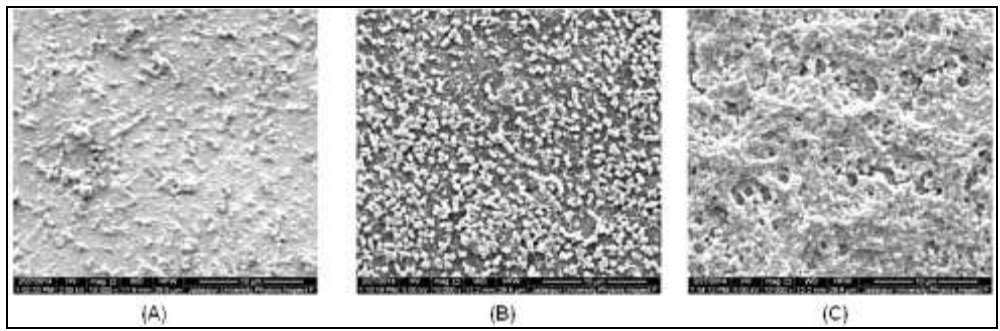

Figure 4. FESEM photographs of ten day old biofilms of P aeruginosa on HDPE surface

(A) On the bare surface, (B) with BSA adsorbed for 9hrs, (C) with BSA adsorbed for 24 hrs.